\documentclass[preprint,aps,prl,amsfonts,letterpaper]{revtex4-1}

\usepackage{CJK} 
\usepackage{graphicx, multirow, amsfonts, amsmath, bbm}
\usepackage{hyperref}
\usepackage[usenames,dvipsnames]{color}

\usepackage[final]{changes} %
\definechangesauthor[name={FZ}, color=blue]{FZ}
\definechangesauthor[name={VO}, color=teal]{VO}

\setremarkmarkup{[#1 : #2]}

\DeclareMathOperator{\Tr}{tr}
\DeclareMathOperator{\diag}{diag}

\begin{document}
	\begin{CJK*}{UTF8}{bsmi}  
		\title{A unified treatment of derivative discontinuity, delocalization and static correlation effects in density functional calculations}
		\author{Fei Zhou(周非)}
		\email{zhou6@llnl.gov}
		\affiliation{Physical and Life Sciences Directorate, Lawrence Livermore National Laboratory, Livermore, CA 94550, USA}
		\author{Vidvuds Ozoli\c{n}\v{s}}
		\email{vidvuds.ozolins@yale.edu}
		\affiliation{Department of Applied Physics, Yale University, New Haven, CT 06520, USA} 
		\affiliation{Yale Energy Sciences Institute, Yale University, West Haven, CT 06516, USA} 
		\date{\today} 
		\begin{abstract}
			We propose a method that incorporates explicit derivative discontinuity of the total energy with respect to the number of electrons and treats both delocalization and static correlation effects in density functional calculations. Our approach is motivated by the exact behavior of the ground state total energy of electrons and involves minimization of the exchange-correlation energy with respect to the Fock space density matrix. The resulting density matrix minimization (DMM) model is simple to implement and can be solved uniquely and efficiently. In a case study of KCuF$_3$, a prototypical Mott-insulator with strong correlation, LDA+DMM correctly reproduced the Mott-Hubbard gap, magnetic ordering and Jahn-Teller distortion.
		\end{abstract}
		\maketitle
	\end{CJK*}

	\begin{figure*}[!hbtp]
		\includegraphics[width=0.215\linewidth]{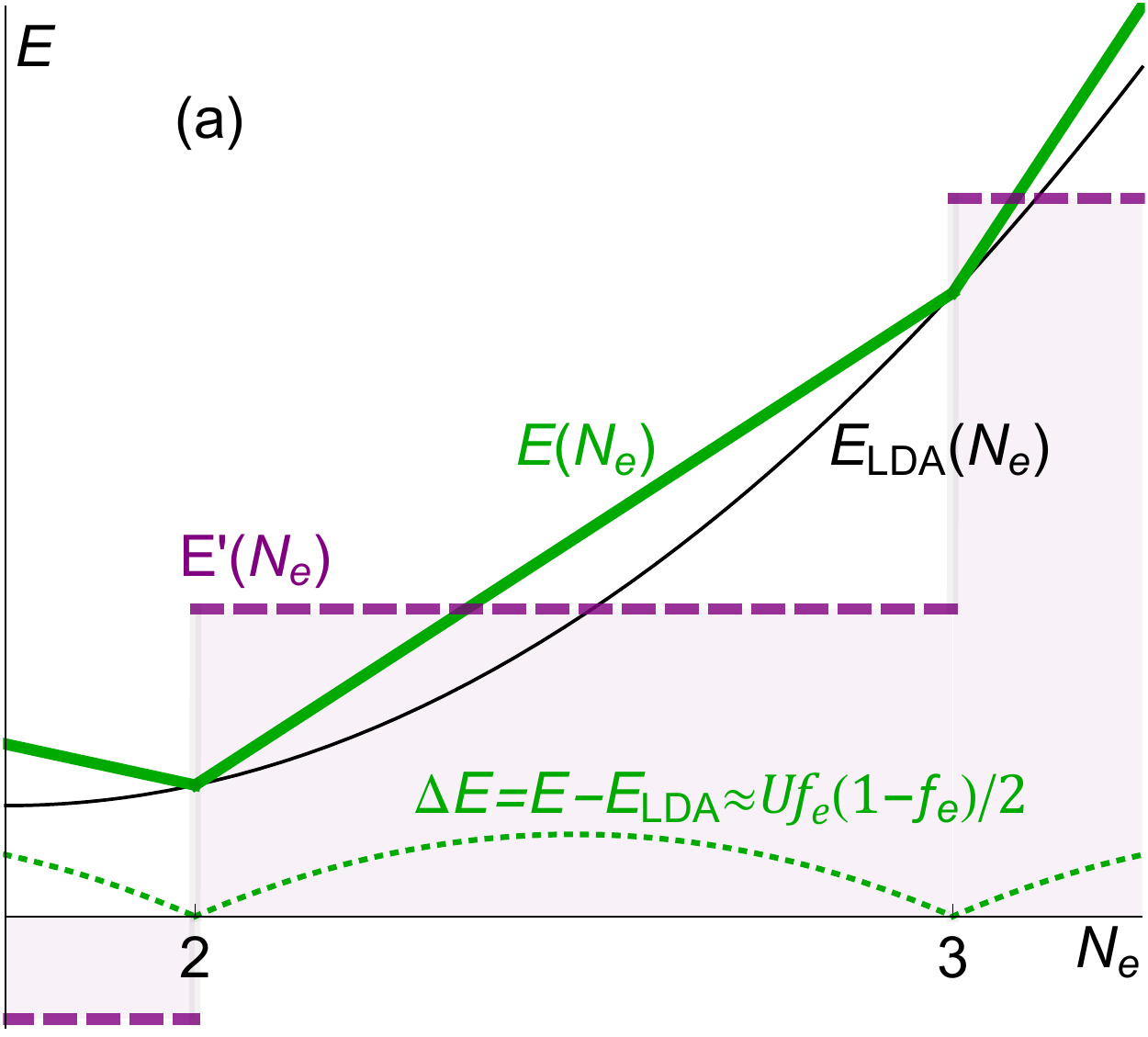}
		\includegraphics[width=0.77\linewidth]{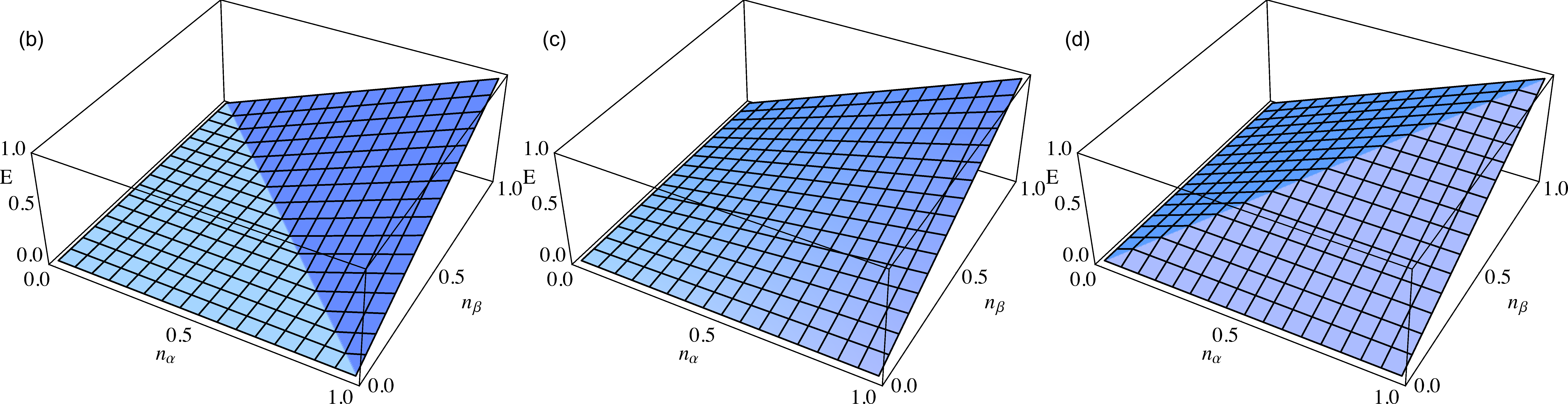}
		\caption{(a) Schematic energy vs.\ $N_\mathrm{e}$, and comparison of $s$-electron Coulomb energy $E_{\mathrm{ee}}$ with $U=1$ and $\mathbf{n}=\diag (n_\alpha, n_\beta) 
			$ minimized by (b) Eq.~\ref{eq:SDP}, (c) mean-field $ n_\alpha n_\beta U$, and (d) the idempotent constraint of Eq.~\ref{eq:pure-state}. 
		}
		\label{fig:MCY} 
	\end{figure*} 
	
Despite its enormous success in computational physics, chemistry and materials science, density functional theory (DFT) with the local density and generalized gradient approximations (LDA, GGA) to the exchange-correlation (xc) is still plagued by major setbacks in strongly correlated systems
\cite{anisimov2010electronic, Cohen2012CR289}. 
The fundamental problems are well known: the delocalization error and the static (non-dynamic) correlation error, arising from self-interaction and multiple competing reference wavefunctions, respectively, are present in not only open-shell $d$/$f$-systems, but main group elements near the atomic limit, e.g.\ bond breaking \cite{Cohen2008S792, Cohen2012CR289}. ``Beyond DFT'' approaches, inspired by  progress in strong correlation physics, e.g.\ on the Hubbard model, have been developed with great successes, including
LDA+$U$ \cite{Anisimov1991PRB943}, 
LDA plus the dynamic mean-field theory (LDA+DMFT) \cite{Georges1996RMP13, Kotliar2006RMP865}, 
and LDA plus the Gutzwiller approximation (LDA+GA) \cite{Ho2008PRB073101, Deng2008EL37008,  *Deng2009PRB075114}. %

The correlation problem of LDA/GGA is best seen by revisiting the foundations of DFT, i.e.,\ the behavior of the exact ground-state total energy $E$. First, $E$ should be piecewise linear in the number of electrons $N_\mathrm{e}$, with discontinuity in $dE/dN_\mathrm{e}$  at integer $N_\mathrm{e}$  \cite{Perdew1982PRL1691, Sham1983PRL1888}. This famous \textit{derivative discontinuity} \cite{Perdew1982PRL1691, Perdew1983PRL1884, Sham1983PRL1888, Mori-Sanchez2008PRL146401} is absent in LDA/GGA due to the delocalization/self-interaction error, 
leading to underestimated charge localization and band gaps at integer $N_\mathrm{e}$ (see Fig.~\ref{fig:MCY}a). In DFT, the band gap is given by the Kohn-Sham (KS) eigenvalue gap plus an {\em explicit} discontinuity from the xc functional \cite{Perdew1983PRL1884,Mori-Sanchez2008PRL146401}:
\begin{align} \label{eq:gap}
	E_\mathrm{g} =  \epsilon_{N_\mathrm{e}+1}^{\mathrm{KS}} - \epsilon_{N_\mathrm{e}}^{\mathrm{KS}} + \mathcal{D}_\mathrm{xc}.
\end{align}
This means that even if the exact KS gap were known, there is still a missing  discontinuity $\mathcal{D}_\mathrm{xc}$, which is particularly important for correct description of Mott insulators \cite{Perdew1983PRL1884, Mori-Sanchez2008PRL146401}. However,  $\mathcal{D}_\mathrm{xc}$ is missing in approximate xc functionals.
Secondly, Mori-S\'anchez, Cohen and Yang investigated another dimension, the spin polarization $n_\alpha - n_\beta$, and found that the ground state energy should remain constant with respect to fractional $n_\alpha - n_\beta$ due to static correlation \cite{Cohen2008JCP121104,Mori-Sanchez2009PRL066403}. For example, a spin-polarized hydrogen atom is degenerate with a non-polarized one, considering that the ground state of a pair of separated hydrogen atoms (1,2) is dominated by the correlated two-electron wavefunction $(|1\alpha 2 \beta \rangle - |1\beta 2 \alpha \rangle )/\sqrt{2}$ ($\alpha,\beta=\uparrow,\downarrow$) with vanishing electron-electron repulsion \cite{Cohen2008JCP121104}.

In this letter, we propose a remedy, LDA plus density matrix minimization (LDA+DMM). Built from the beginning with the above conditions for the exact ground state total energy in mind, our approach offers significant quantitative improvement in total energies over LDA and LDA+$U$ for strongly correlated systems. Electronic structure predictions of LDA+DMM
overcome qualitatively the failures of LDA and LDA+$U$ in Mott insulators at a modest computational cost. In the following, the DMM model is presented, followed by a case study on a prototypical Mott insulator, KCuF$_3$.

Our starting point is an isolated atom with open $l$-shell (fractional $N_\text{e}$) and spin-orbitals $| i \rangle \equiv | lm\sigma \rangle$ as  one-body basis designated by composite index $i=1,\dots,4l+2$. With the assumption of identical radial wavefunctions, the kinetic and external potential energies are simply linear in $N_\text{e}$. This means that the above conditions for total energy apply to the on-site electron-electron repulsion $E_\text{ee}$. As shown in Fig.~\ref{fig:MCY}b for isolated $s$-electrons 
(see also Fig.~1 of Ref.~\onlinecite{Mori-Sanchez2008PRL146401}), $E_\mathrm{ee}$ is linear in $N_\mathrm{e}= n_\alpha + n_\beta$ and constant in $n_\alpha - n_\beta$, while neither local/hybrid functionals \cite{Mori-Sanchez2009PRL066403} nor LDA+$U$ (Fig.~\ref{fig:MCY}c) follow these conditions. 

We discuss the many-body physics of correlated $l$-electrons in the Fock space  with $2^{4l+2}$ basis functions $\Phi_I$ ($I=1,\dots,2^{4l+2}$) chosen as $C_{4l+2}^N$ Slater determinants in $N$-electron subspaces ($0\leq N \leq 4l+2$). 
The Coulomb operator $\hat{V}_\text{ee}$  becomes a block-diagonal matrix \begin{align*}
\mathbb{V}_\text{ee}  =  
\text{diag}\left(V_\text{ee}^{(0)} , V_\text{ee}^{(1)} , \cdots , V_\text{ee}^{(4l+2)}\right),
\end{align*}
where $V_\text{ee}^{(N)}$ is the $C_{4l+2}^N \times C_{4l+2}^N$ matrix of Coulomb repulsion in the $N$-body subspace.
While a {\em pure} quantum sate of the Fock-space is an appropriate description for isolated atoms \added{completely cut off from the outside world},
a partially filled shell with fractional $N_\text{e}$ is implicitly part of a larger environment, and is in a {\em mixed} quantum state, such as used in Perdew's original treatment on fractional $N_\text{e}$ in DFT \cite{Perdew1982PRL1691}. We therefore choose the density matrix as the fundamental variable describing the electronic correlations \cite{RevModPhys.29.74}. Mathematically, it is also a block-diagonal matrix written as 
\begin{align*}
\mathbb{D}  =  
\text{diag}\left(D^{(0)} , D^{(1)} , \cdots , D^{(4l+2)}\right),
\end{align*}
where $D^{(N)}$ is the $N$-body density matrix, and $\Tr D^{(N)}$ designates the probability of finding the quantum state with $N$ electrons. Since the eigenvalues of the density matrix have the physical meaning of probabilities,  $\mathbb{D}$ is positive semidefinite ($\mathbb{D} \succeq 0$) \cite{RevModPhys.29.74}.
An observable such as Coulomb repulsion is given by $E_\text{ee}[\mathbb{D}] = \Tr \mathbb{V}_\text{ee} \mathbb{D}$. In the context of DFT calculations, the correlated subspace of the partially filled $l$-shell is linked to the KS wavefunctions via the on-site occupancy matrix (OOM) $\mathbf{n}$. For the Fock-space density matrix, it is given by expectancy $\Tr \mathbb{N}_{ij} \mathbb{D}$  of the projection operator $\hat{c}^\dagger_{i} \hat{c}_j$, where $ \mathbb{N}_{ij}  $ is the matrix for $\hat{c}^\dagger_{i} \hat{c}_j$. For KS-DFT, $\mathbf{n}$ is the  projected from  Kohn-Sham orbitals: $ n_{ij} = \langle \Psi_\text{KS} |\hat{c}^\dagger_{i} \hat{c}_j | \Psi_\text{KS} \rangle =  \sum_{nk} f_{nk} \langle  \psi_{nk} | i \rangle \langle j |\psi_{nk} \rangle$. We require that they match: $\Tr \mathbb{N}_{ij} \mathbb{D} = n_{ij}$.

Like in LDA+$U$, the LDA+DMM total energy is given by the DFT total energy plus the electron-electron interaction, minus the double counting (dc) term:
\begin{eqnarray}
E_{\mathrm{LDA+DMM}}&= &
  E_{\mathrm{LDA}}   + \sum_{S} \left( E_{\mathrm{ee}}^{\mathrm{DMM}} - E_\text{dc} \right), \label{eq:LDA+DMM}
\end{eqnarray}
where the summation is over correlated sites $S$. The DMM energy is minimized over the density matrix
\begin{align}  \label{eq:SDP}
&E_{\mathrm{ee}}^{\mathrm{DMM}}[\mathbf{n}] = \min_{\mathbb{D}} \Tr\mathbb{V}_\text{ee} \mathbb{D}\\
\text{s.t.}&\ \mathbb{D} \succeq 0,\ \Tr \mathbb{D} =1,\ \Tr \mathbb{N}_{ij}  \mathbb{D} =n_{ij} \nonumber
\end{align}
under two constraints: a) positive semidefinite and normalized $\mathbb{D}$, and b) matching of occupancy $\mathbf{n}$.  
The total energy (\ref{eq:LDA+DMM}) then is minimized with respect to KS orbitals while the DFT effective potential contains a contribution arising from the Lagrange multipliers
$$
V_{ij}^\text{DMM} = \partial E_{\mathrm{ee}}^\text{DMM}/ \partial n_{ij}
$$
corresponding to the $\Tr \mathbb{D} \mathbb{N}_{ij}=n_{ij}$ constraints.

A key advantage of Eq.~(\ref{eq:SDP}) is that it constitutes a semidefinite programming problem (SDP)  \cite{Vandenberghe1996SR49, Todd2001AN515}, a well-known convex optimization problem \cite{Boyd-Vandenberghe} that can be solved \textit{uniquely} and \textit{efficiently}  with  numerical algorithms \cite{wolkowicz2000handbook, Monteiro2003MP209, Borchers1999OMS613}, which are capable of minimizing Eq.~(\ref{eq:SDP}) within few seconds for $d$-electrons. The optimized dual variables of SDP \cite{Vandenberghe1996SR49, Todd2001AN515}  %
yield $V_{ij}^\text{DMM}$.

To illustrate our approach, we consider the simple case of partially filled $s$-shell ($l=0$). The Coulomb matrix becomes
$
\mathbb{V}_\text{ee} = \text{diag}\left(0 , 0, 0, U\right).
$
The density matrix is in general $\mathbb{D} =\text{diag}\left(P^{(0)} , \left[ \begin{array}{cc}
	P^{(1)}_{\alpha \alpha},P^{(1)}_{\alpha \beta} \\ P^{(1)}_{\beta \alpha},P^{(1)}_{\beta \beta}
\end{array}\right] , P^{(2)}\right)$. Now reconsider the previous example of separated H$_2$ with two-body wavefunction $(|1\alpha 2 \beta \rangle - |1\beta 2 \alpha \rangle )/\sqrt{2}$.  From the perspective of one hydrogen atom, projection to this site yields the on-site density matrix $\mathbb{D}= \diag(0,\frac12, \frac12, 0) $ (i.e.\ $ (|\alpha \rangle \langle \alpha|  + |\beta \rangle \langle \beta| )/2$), and occupancy $\mathbf{n}= \diag(\frac12, \frac12)$. It is easy to check in this simple case that eq.~(\ref{eq:SDP}) reduces to a linear programming problem and the above $\mathbb{D}$ indeed corresponds to a minimized $E_\text{ee}^\text{DMM}=0$, which satisfies the charge-linearity/spin-constancy conditions for $s$-electrons (Fig.~\ref{fig:MCY}b). In general, $E_{\mathrm{ee}}^{\mathrm{DMM}}[\diag(n_\alpha, n_\beta)]$ of Eq.~(\ref{eq:SDP}) meets these conditions on the whole $(n_\alpha, n_\beta)$ plane, as shown in Fig.~\ref{fig:MCY}b. In contrast, the mean-field approximation  $E_{\mathrm{ee}} \approx n_\alpha n_\beta U$ (Fig.~\ref{fig:MCY}c) of LDA+$U$ deviates from the exact $E_{\mathrm{ee}}$ except at integer occupancy \footnote{See Supplemental Material for detailed examples.}. 
	Fig.~\ref{fig:MCY}d further elucidates the physical origin of the exact behavior: static correlation due to the presence of alike atoms. If one artificially turns off such correlation by replacing the many-body density matrix $\mathbb{D}$ for a \textit{mixed} state with a Fock-space \textit{pure} state, and accordingly replacing the mixed-state average $\langle A \rangle _\mathbb{D}=\Tr \mathbb{D} \mathbb{A} $ of operator $\hat{A}$ in Eq.~(\ref{eq:SDP}) with the pure state expectation $\langle A \rangle _\Psi = \langle \Psi |  \hat{A} | \Psi \rangle$, one switches to a pure state picture: 
	\begin{eqnarray}
	E_{\mathrm{ee}}^{\mathrm{ps}}[n] &=& \min_{|\Psi| =1, \langle \hat c_i ^\dagger \hat c _j \rangle _\Psi=n_{ij}}  \langle \mathbb{V}_{\mathrm{ee}} \rangle _\Psi . \label{eq:pure-state}
	\end{eqnarray}
	This is equivalent to restricting the search space of $\mathbb{D}$ to idempotent matrices. The pure state formalism is, again, correct only at integer occupancy (Fig.~\ref{fig:MCY}d). Otherwise, the difference from Fig.~\ref{fig:MCY}b is striking. At a given $N_\mathrm{e}$, the pure state formalism strongly favors the maximum amount of spin polarization, in dramatic violation of the spin-constancy condition.
	For example, $\mathbf{n}=\diag(\frac12, \frac12)$ corresponds to the $N_\mathrm{e} =1$ mixed state $D^{(1)}= \frac12 \sum_{\sigma} | \sigma \rangle \langle \sigma |$ with $E_{\mathrm{ee}}= 0$, or to the pure state $\Psi=(| \rangle + |\alpha,\beta  \rangle)/\sqrt2$ with $E_{\mathrm{ee}}= U/2$. 
	This justifies our choice of the mixed-state density matrix $\mathbb{D}$ as the basic variational variable \cite{Note1}. 

\begin{figure}[htbp]
		\includegraphics[width=0.99\linewidth]{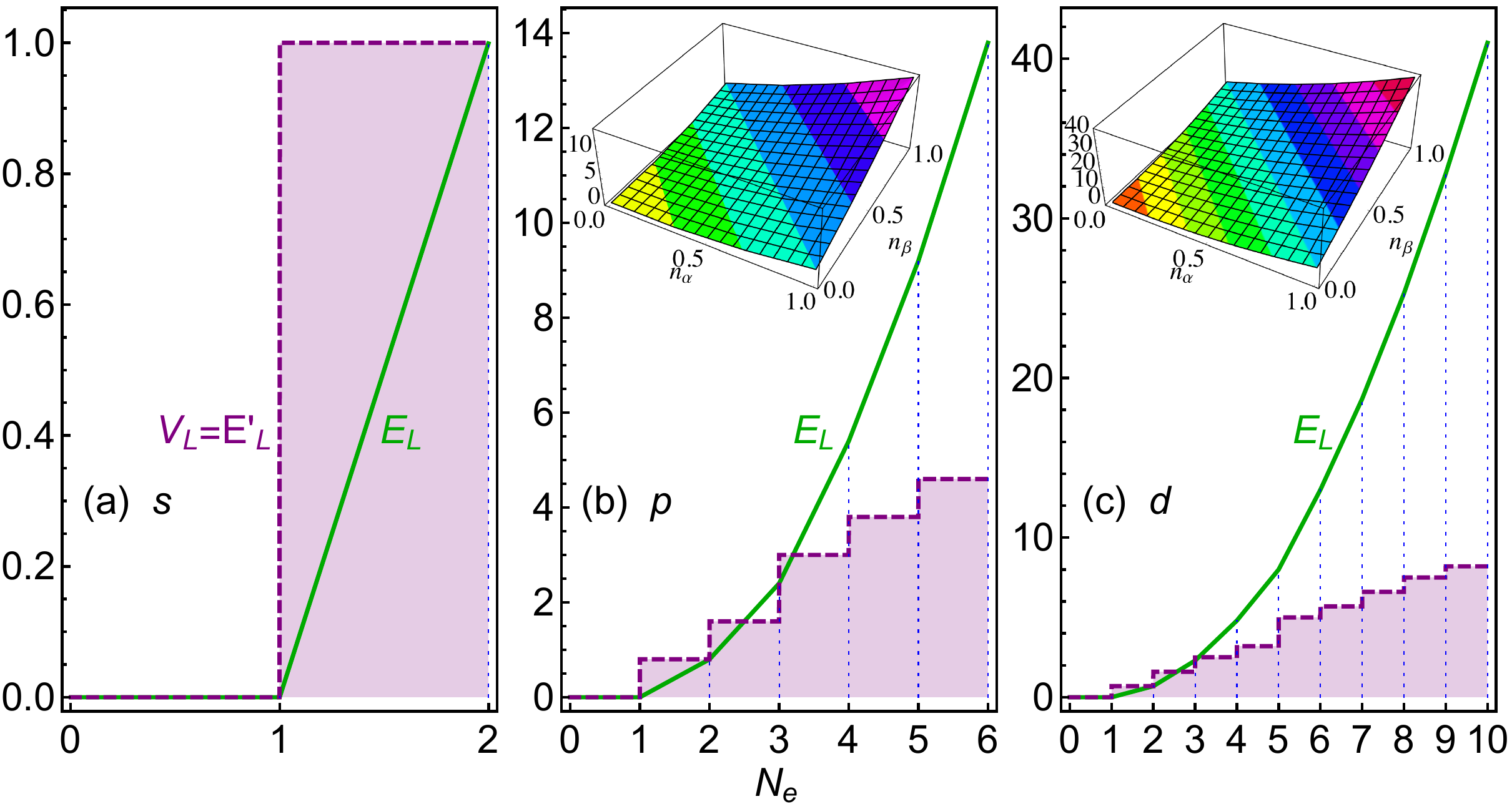}
		\caption{\label{fig:discontinuity}
			With $\mathbf{n}=\diag(n_\alpha, \dots, n_\alpha, n_\beta, \dots, n_\beta)$, $U$=1, $J$=0.2, energy $E_L(N_{\mathrm{e}})=E_{\mathrm{ee}}^{\mathrm{DMM}}(\mathbf{n})$ (solid line) and potential $V_{L}=dE_L(N_{\mathrm{e}})/dN_{\mathrm{e}}$ (purple dashed line) vs.\  $N_{\mathrm{e}}$ for $l=0 \mbox{--}2$, respectively. Inset: $E_{\mathrm{ee}}^{\mathrm{DMM}}$ vs.\ $n_\alpha$, $n_\beta$.
		}
	\end{figure}

	DMM is applicable beyond $s$-electrons. Fig.~\ref{fig:discontinuity} shows $E_{\mathrm{ee}}^\mathrm{DMM}$ vs.\ $(n_\alpha, n_\beta)$ for $l=1,2$ with $n_{ij}= n_\sigma \delta_{ij}$ spherical in each spin channel. Similar to $l=0$, Eq.~(\ref{eq:SDP}) satisfies the conditions for fractional charge and spin.
	Fig.~\ref{fig:discontinuity}a-c shows the piecewise straight line $E_L(N_{\mathrm{e}})= E_{\mathrm{ee}}^\mathrm{DMM}$, as expected of the ground state energy of fractional $N_\mathrm{e}$ \cite{Perdew1982PRL1691}. The calculated potential $V_{ij} = V_L \delta_{ij} $ is spherical,  spin-independent, and  piecewise constant (dashed lines). An \textit{explicit} derivative discontinuity $\mathcal{D}_\mathrm{xc} =   V_{\mathrm{ave}}(z_\mathrm{e}+\delta) -V_{\mathrm{ave}}(z_\mathrm{e}-\delta) $ at integer $z_\mathrm{e}$  is recovered:
	\begin{eqnarray} \label{eq:DV}
	\lim_{N_\mathrm{e} \rightarrow z_\mathrm{e}^+}V_{ij} - \lim_{N_\mathrm{e} \rightarrow z_\mathrm{e}^-}V_{ij} = \mathcal{D}_\mathrm{xc} \delta_{ij} = \left[U-x(\mathbf{n}) J \right] \delta_{ij} , %
	\end{eqnarray}
	where $V_{\mathrm{ave}} = \Tr \mathbf{V}/(4l+2)$ is the average potential and $x$ is an coefficient for $J$.

\begin{figure}[hbtp] 
		\includegraphics[width=0.494 \linewidth]{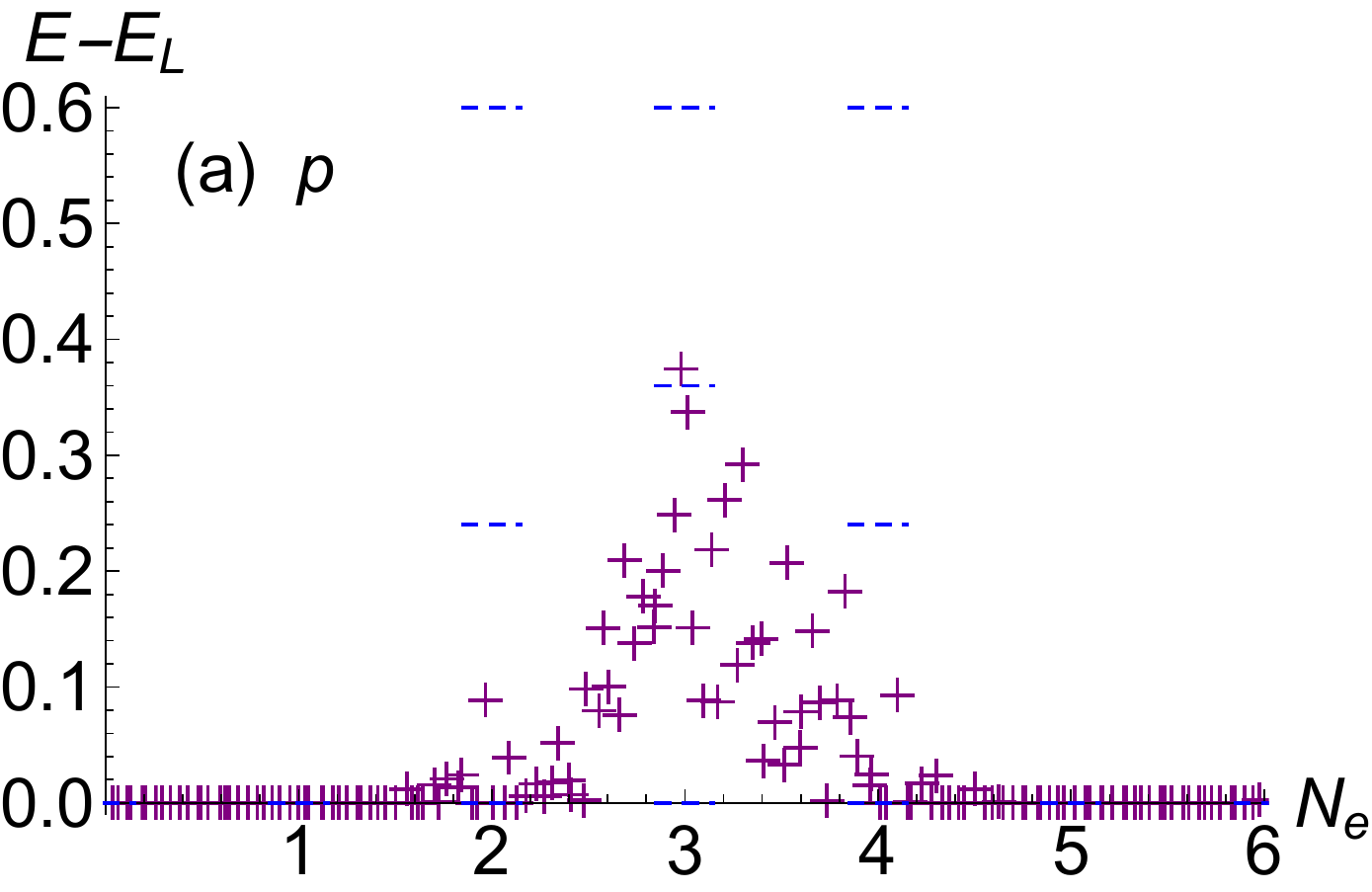} 
		\includegraphics[width=0.494 \linewidth]{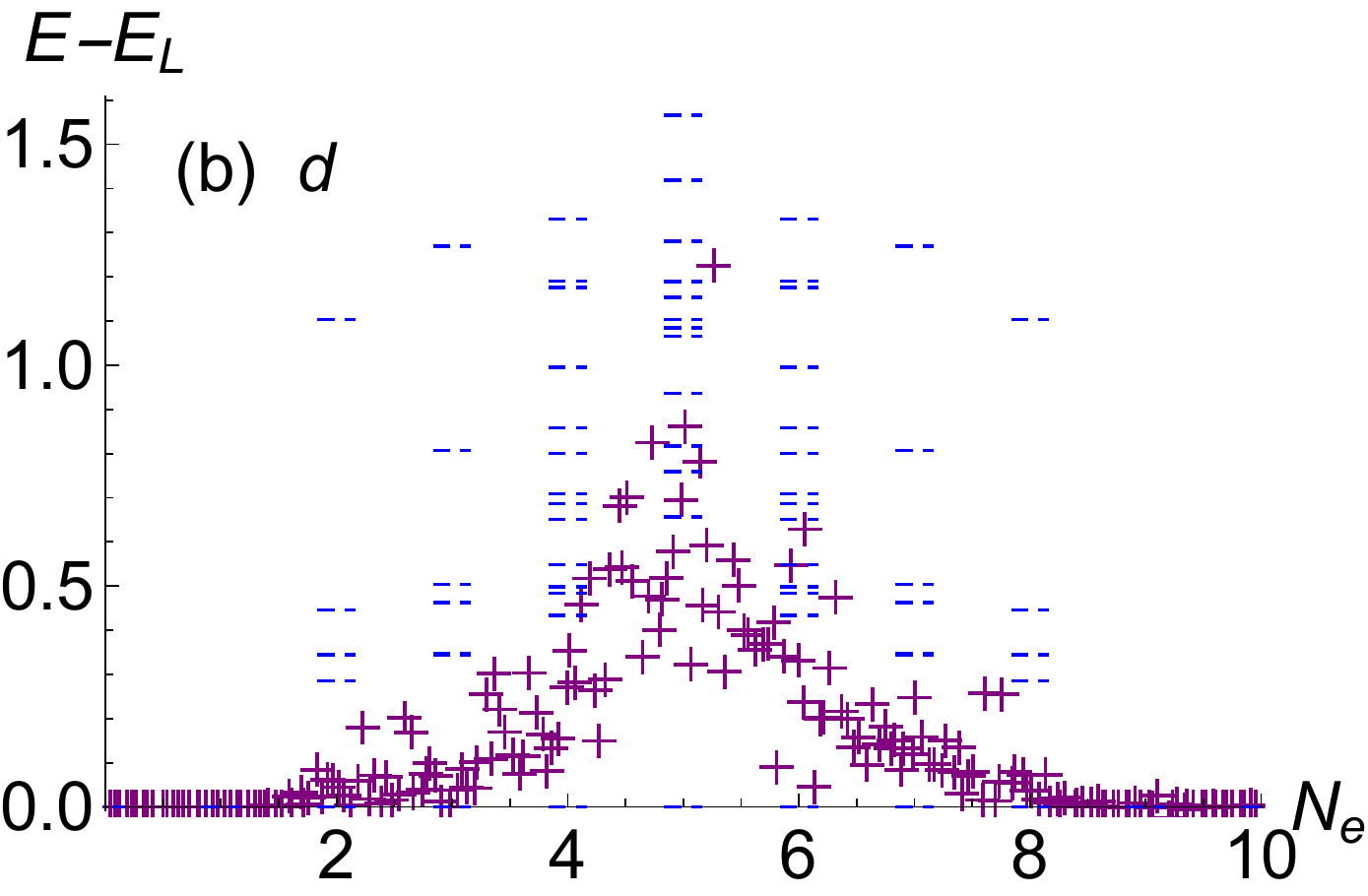} 
		\includegraphics[width=0.68 \linewidth]{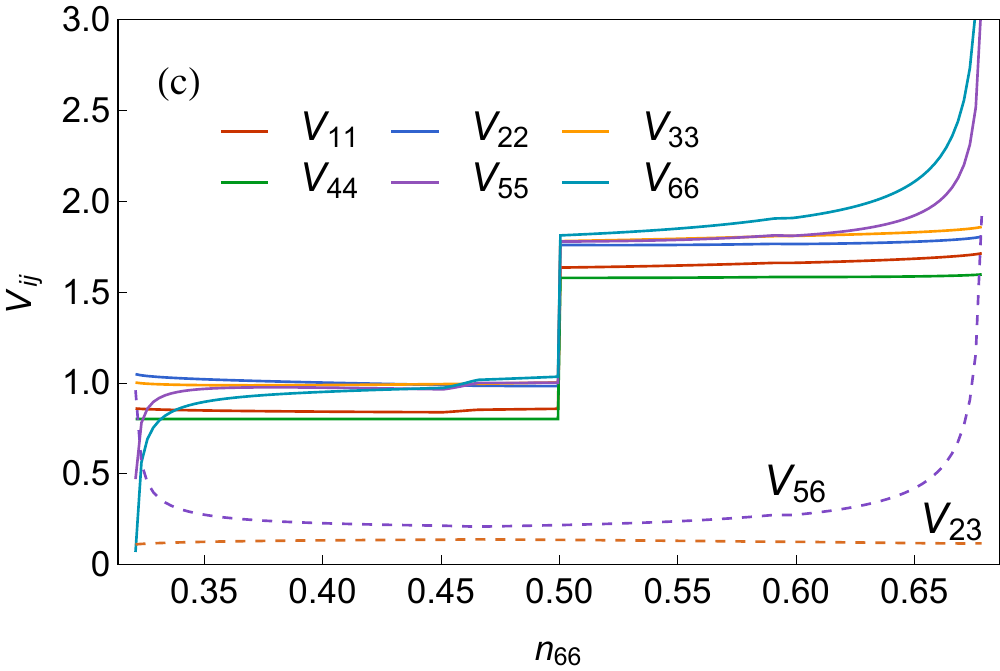} 
		\caption{With $U$=1, $J$=0.2, $E_{\mathrm{ee}}^{\mathrm{DMM}}$  of (a) $p$ and (b) $d$-electrons with random diagonal $\mathbf{n}$ vs.\  $N_{\mathrm{e}}$, with eigen-energies of $ \mathbb{V}_{\mathrm{ee}}^{N}$ shown as dashed lines; %
			and (c)  $V_{ij}$ for $l=1$, $\mathbf{n}=\diag(0.1,0.2,0.3, 0.4, \left[\begin{array}{cc}
			0.5 & 0.4 \\ 
			0.4 & n_{66}
			\end{array} \right])$ vs.\ $n_{66}$.} \label{fig:electron-rand-energy} 
	\end{figure}

	All the above examples fall on the ground state line $E_L$, suggesting that the supplied OOMs $\mathbf{n}$ can be interpolated by those of atomic ground states.  We call $\mathbf{n}$ \textit{linear representable} if $E_{\mathrm{ee}}^\mathrm{DMM}[\mathbf{n}] = E_L(N_\mathrm{e})$.
	However, these examples are the exception rather than the norm.  For $l>0$, it is generally not possible to interpolate an arbitrary $\mathbf{n}$ with ground state OOMs alone, i.e.\ $E_{\mathrm{ee}}^\mathrm{DMM}[\mathbf{n}] > E_L(N_\mathrm{e})$. 
	This can be seen from Fig.~\ref{fig:electron-rand-energy} showing the energy of random diagonal OOMs. For $l=1$ and $1<N_{\mathrm{e}}<5$ (Fig.~\ref{fig:electron-rand-energy}a), the energy of a large number of OOMs is  above $E_L$, i.e.\ not linear representable. The same holds for $l=2$ and $1<N_{\mathrm{e}}<9$ (Fig.~\ref{fig:electron-rand-energy}b). For a $p^2$ example of $\mathbf{n}=\diag (1,0,0,1,0,0)$, which violates Hund's first rule, $E_{\mathrm{ee}}^\mathrm{DMM}[\mathbf{n}]= U+0.8 J>E_L(2)=U-J$. Exceptions include the trivial case of $N_{\mathrm{e}} \leq 1$ or $N_{\mathrm{e}}\geq M-1$ (and hence any $s$-system), where the degeneracy of a single electron/hole means no excited states, as well as special $n$ such as the previous spin-spherical $n_{ij}= n_\sigma \delta_{ij}$.
	
	Note that in the above linear representable examples in Fig.~\ref{fig:discontinuity}, $\mathbf{V} = V_L \mathbbm{1} $ was found spherical and spin-independent. 
	It can be shown that this is true for any linear representable OOM, not considering discontinuity at integer $N_\mathrm{e}$.  The physical meaning of $\mathbf{V} = V_L \mathbbm{1} $ is remarkable: essentially, the ground state can be reproduced with a spin-independent potential, as it should be in true DFT. This is not the case for LDA+$U$.
	Now the DMM energy $E_{\mathrm{ee}}^\text{DMM}$ can be understood as the ground state $E_{L}$ plus a penalty for departure from linear representability. Similarly $\mathbf{V}$ is composed of a scalar part $V_{L}$, plus an aspherical contribution driving the OOM towards linear representability and compatibility with Hund's rules. 
	For example, the dependence of $\mathbf{V}$ on $n_{66}$ is shown in Fig.~\ref{fig:electron-rand-energy}c, with a uniform derivative discontinuity $\mathcal{D}_\mathrm{xc}$ as in Eq.~(\ref{eq:DV}) at $z_{\mathrm{e}}=2$, and singularity in $V_{55}$, $V_{56}$, $V_{66}$ when an eigenvalue of $\mathbf{n}$ approaches 0 or 1.

	\begin{figure*}[hbtp] 
		\includegraphics[width=0.34 \linewidth]{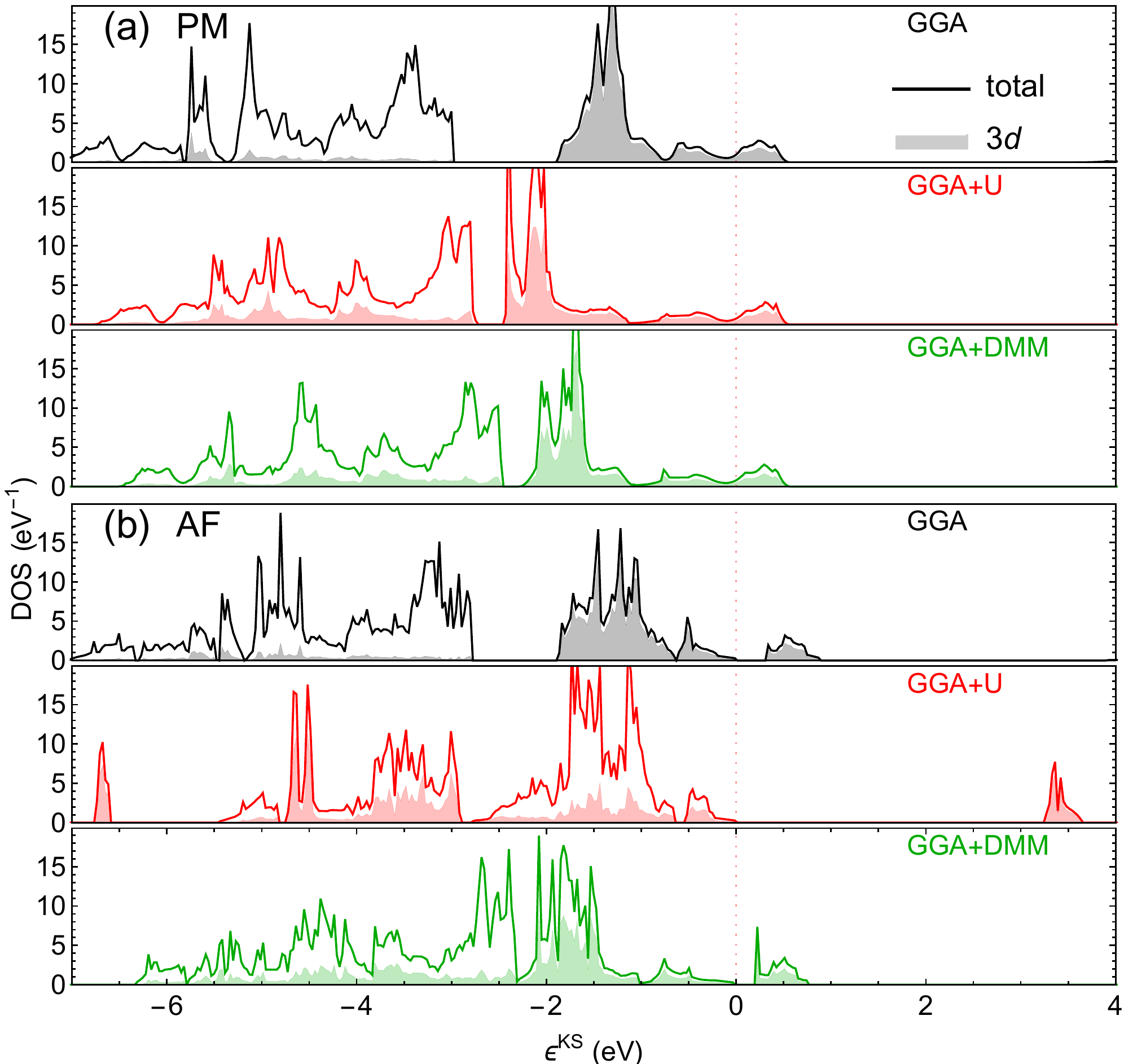} 
		\includegraphics[width=0.29 \linewidth]{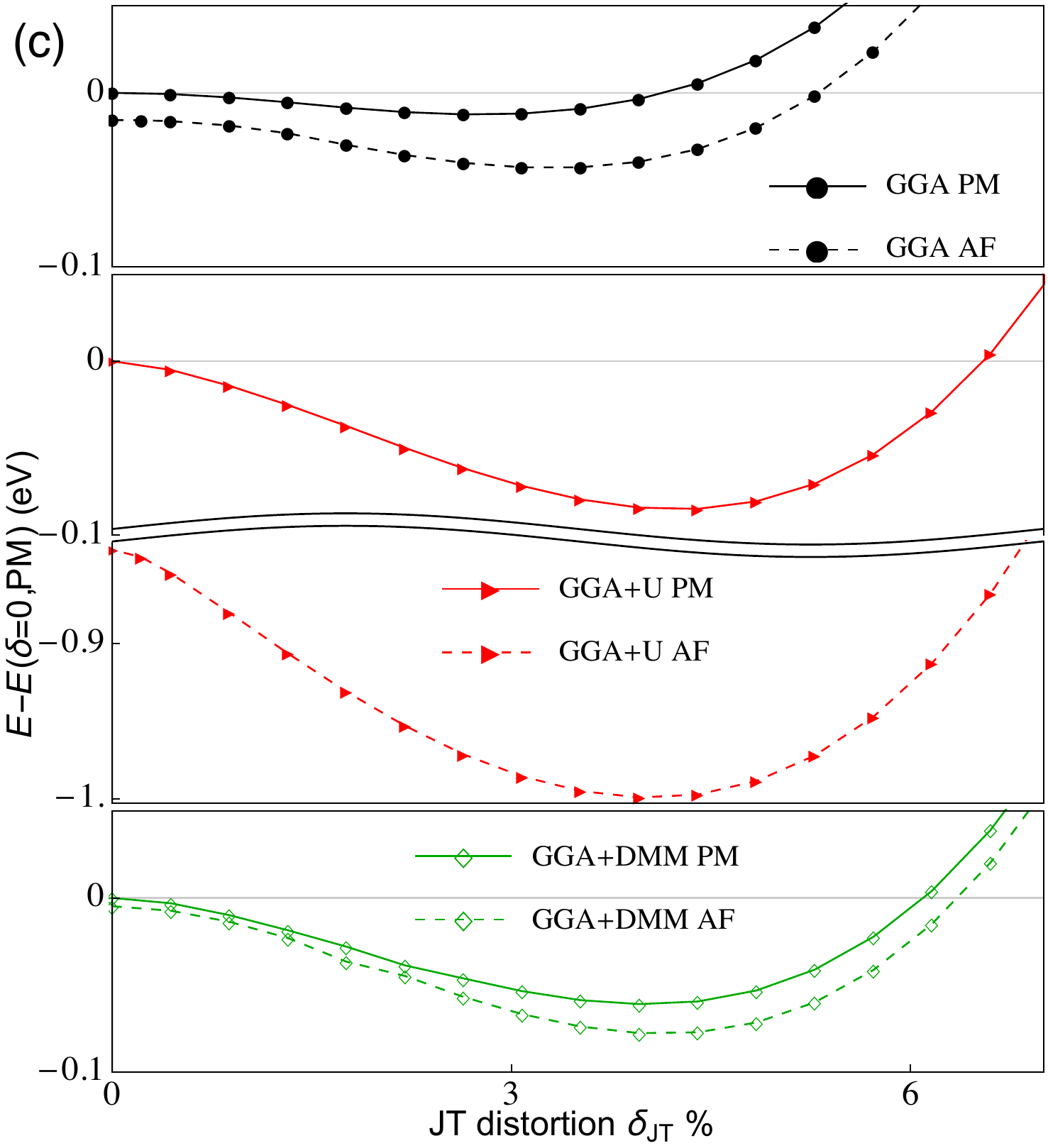} 
		\includegraphics[width=0.33 \linewidth]{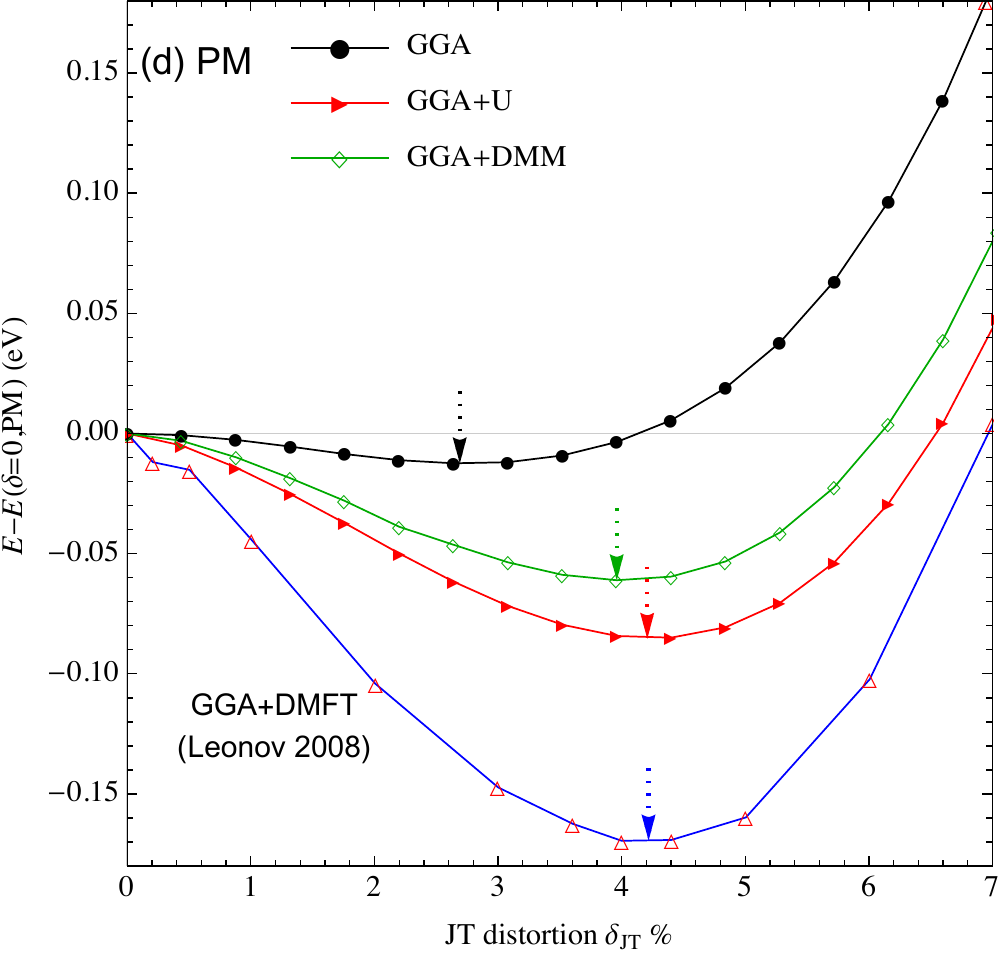} 
		\caption{For KCuF$_3$: density of states in the (a) paramagnetic and (b) antiferromagnetic phase, and energy per formula unit versus Jahn-Teller distortion (c) in the two phases and (d) in PM compared to GGA+DMFT results from Ref.~\onlinecite{Leonov2008PRL096405}. The optimized distortion is pointed out by arrows.} \label{fig:KCuF3} 
	\end{figure*}
	Next, we present GGA+DMM calculations for KCuF$_3$, a prototypical Mott insulator.
	Correct reproduction of Mott-Hubbard gap, orbital ordering and Jahn-Teller distortion in the antiferromagnetic (AF) phase of KCuF$_3$ was one of the early achievements of LDA+$U$  \cite{Liechtenstein1995PRBR5467}. However, the paramagnetic (PM) phase was beyond the capabilities of LDA+$U$ due to strong static correlation.  Leonov and coworkers \cite{Leonov2008PRL096405} studied the PM phase with DFT+DMFT calculations, and successfully reproduced the Mott band gap and the observed Jahn-Teller distortion (4.4\% \cite{Buttner1990ACB131}).

We adopt the double counting (dc) scheme of our previous work \cite{Zhou2009PRB125127, *Zhou2011PRB085106, *Zhou2012PRB075124} by separating the dc energy into the Hartree energy and the xc contribution in order to avoid aspherical self-interaction errors:
	\begin{eqnarray}
	E_{\mathrm{dc}}[\mathbf{n}] &=& E_{\mathrm{H}} [\mathbf{n}] + E_{\mathrm{dc}}^{\mathrm{xc}} [\mathbf{n}],  \\
	E_{\mathrm{H}} &=& \frac12 \sum_{ \{m\} }  \langle m, m'' | \hat{v}_\mathrm{ee} | m' ,m'''  \rangle n_{m m'} n_{m'' m'''} \label{eq:Hartree}, \\
	E_{\mathrm{dc}}^{\mathrm{xc}} &=& - \frac12 [ U N_\mathrm{e} +  J  N_\mathrm{e} (N_\mathrm{e}-2)/4 ].  \label{eq:dcX}  
	\end{eqnarray}
	This allows one to correct the xc energy, not the Hartree term, which is exact by definition in DFT and does not need a dc approximation \cite{Zhou2009PRB125127, *Zhou2011PRB085106, *Zhou2012PRB075124}.
	$E_{\mathrm{dc}}^{\mathrm{xc}}$ in Eq.~(\ref{eq:dcX}) is the one used in Ref.~\onlinecite{Leonov2008PRL096405}.
	Accordingly the correction potential is
	\begin{eqnarray}
	\Delta V_{ ij} &=& V_{ij} - \partial E_{\mathrm{H}}/\partial n_{ij} - \partial E_{\mathrm{dc}}^{\mathrm{xc}} / \partial n_{ij}. \label{eq:xcpot}
	\end{eqnarray}

	Before delving into numerical details, we point out a qualitative feature of LDA+DMM. In the so-called spherically averaged or $J=0$ limit, Eq.~(\ref{eq:SDP}) is simply linear interpolation of $U z_\mathrm{e}(z_\mathrm{e}-1)/2$ between integers $z_\mathrm{e}$, and Eq.~(\ref{eq:LDA+DMM}) becomes
	\begin{align}
	E_{\mathrm{LDA+DMM}}(J=0) = E_{\mathrm{LDA}} + \sum_S U f_\mathrm{e}(1-f_\mathrm{e})/2,
	\end{align}
	where $0 \leq f_\mathrm{e} <1$  and $N_\mathrm{e}=z_\mathrm{e}+f_\mathrm{e}$. This is exactly the well-known self-interaction correction for fractional number of electrons (Fig.~\ref{fig:MCY}a). In this simplified picture, LDA+DMM corrects the convexity of LDA, in contrast to the mean-field LDA+$U$, which corrects for occupancy of each orbital with $\sum_i U n_i(1-n_i)/2$, the root cause of its multiple minima problems. When the $U$ parameter is large enough, $N_\mathrm{e}$ is pinned to $z_\mathrm{e}$ accompanied by an abrupt derivative discontinuity $\mathcal{D}_\mathrm{xc}=U$ from Eq.~(\ref{eq:DV}). Note that appropriate, finite $J$ values are still important for quantitative accuracy.
	
	This is indeed observed in our GGA+DMM calculations for KCuF$_3$ ($U_\mathrm{critical}$=8.06 eV at $J$=0.9 eV \footnote{see Supplemental Material for details}). Fig.~\ref{fig:KCuF3}ab compares the obtained total and projected density of states (DOS) with GGA and GGA+$U$. GGA predicts metallicity in the PM phase. Both GGA+$U$ and GGA+DMM push occupied 3$d$ states down with no KS gap. The difference is, while the former cannot handle static correlation, GGA+DMM predicts correctly a Mott gap of $\mathcal{D}_\mathrm{xc}  \approx 7$ eV according to Eqs.~(\ref{eq:gap},\ref{eq:DV}). Note that GGA+DMFT predicts a much smaller band gap $\sim 1.5-3.5$ eV \cite{Leonov2008PRL096405}. In the AF phase (Fig.~\ref{fig:KCuF3}b), all methods were able to stabilize antiferromagnetic holes. Both GGA and GGA+DMM predict a tiny KS gap ($\sim 0.3$ eV). The latter again should be augmented by $\mathcal{D}_\mathrm{xc}$. GGA+$U$ predicts a larger KS gap of 3.2 eV. Of the three methods, only GGA+DMM was able to predict a Mott insulator for both magnetic configurations.

	The total energy properties are more interesting. Fig.~\ref{fig:KCuF3}c compares the energy profile  for the PM (solid line) and AF (dashed line) configurations vs.\ the $\delta_\mathrm{JT}$ parameter \cite{Leonov2008PRL096405} for the degree of Jahn-Teller distortion. The reference point was chosen as the paramagnetic, undistorted ($\delta_\mathrm{JT}=0$, space group P4/mmm) structure. All the methods predict correctly the AF ground state. Quantitatively, both GGA and GGA+DMM predict slightly more stable AF state, in qualitative agreement with the low N\'eel temperature of 38 K. In contrast, GGA+$U$ penalizes the PM configuration  too heavily, in overall agreement with previous LDA+$U$ studies \cite{Binggeli2004PRB085117}, due to lack of treatment for static correlation.
	
	Finally, the PM energy profiles are compared in Fig.~\ref{fig:KCuF3}d, together with GGA+DMFT results from Ref.~\onlinecite{Leonov2008PRL096405}. Together with Fig.~\ref{fig:KCuF3}c, one observes that GGA stabilization of JT distortion is much too weak, particularly in the PM phase. Both GGA+$U$ and GGA+DMM predict much larger stabilization energy and amount distortion: $\delta_\mathrm{JT}=$ 4.0\% and 4.2\%, respectively, in good agreement with experimental 4.4\%. GGA+DMFT predicts similar distortion with even stronger stabilization energy than the former two. There is yet no clear experimental data to establish quantitatively the JT distortion energy.

	In conclusion, built with the exact behavior of the ground state total energy of electrons in mind, LDA+DMM offers unified treatment of derivative discontinuity, delocalization errors and static correlation errors in density functional calculations, with clear advantage over LDA and LDA+$U$ for strongly correlated systems. As the first generally applicable method to incorporate explicit derivative discontinuity, LDA+DMM correctly reproduced the Mott-Hubbard gap, even in the presence of strong static correlation, as well as more accurate total energies. 
	The fact that DMM is easy to add in any DFT code implementing LDA+$U$ and that the underlying semidefinite programming problem can be solved uniquely and efficiently, makes it especially attractive.  
	Furthermore, LDA+DMM provides physical insight into the requirements for incorporating the derivative discontinuity and correcting the static correlation and delocalization errors of the current xc functionals. We expect that this method will be a useful tool for future DFT-based studies of strongly correlated materials.

\begin{acknowledgments}
		We acknowledge helpful discussions with L.\ Vandenberghe, B.\ O'Donoghue and B.\ Sadigh. The work of F.Z\ was supported by the Laboratory Directed Research and Development program at Lawrence Livermore National Laboratory and the Critical Materials Institute, an Energy Innovation Hub funded by the U.S. Department of Energy, Office of Energy Efficiency and Renewable Energy, Advanced Manufacturing Office, and performed under the auspices of the U.S. Department of Energy by LLNL under Contract DE-AC52-07NA27344. V.O. was supported by the U.S. Department of Energy, Office of Science, Basic Energy Sciences, under Grant DE-FG02-07ER46433. We acknowledge use of computational resources from the National Energy Research Scientific Computing Center, which is supported by the Office of Science of the U.S. Department of Energy under Contract No. DE-AC02-05CH11231.
\end{acknowledgments}

\end{document}